\documentclass[a4paper]{jpconf}
\usepackage{graphicx}
\usepackage{bm,color}
\usepackage{braket}
\usepackage{amsmath,amssymb}

\begin{document}
\title{Is it possible to study neutrinoless $\beta\beta$ decay
	by measuring double Gamow-Teller transitions?}

\author{Javier Men{\'e}ndez$^1$, Noritaka Shimizu$^2$ and Kentaro Yako$^3$}
\address{Center for Nuclear Study, University of Tokyo, Bunkyo-ku, Hongo, 113-0033, Japan}
\ead{
\\$^1$menendez@cns.s.u-tokyo.ac.jp
\\$^2$shimizu@cns.s.u-tokyo.ac.jp
\\$^3$yako@cns.s.u-tokyo.ac.jp}

\begin{abstract}
Searches of neutrinoless double-beta decay require information on the value
of the nuclear matrix elements that rule the process
to plan and interpret experiments.
At present, however, even the matrix elements obtained
with the most reliable many-body approaches
do not agree to each other better than a factor two or three.
A usual test of the many-body calculations
is the comparison to several nuclear observables,
but so far no nuclear structure property has been found
to show a good correlation to neutrinoless double-beta decay.
Here we propose that double charge-exchange experiments
can offer a very valuable tool to provide insights
on neutrinoless double-beta decay.
Double charge-exchange reactions are being currently performed
in various laboratories worldwide and aim to find the novel nuclear collectivity
given by double Gamow-Teller excitations.
Our results suggest a good linear correlation
between double Gamow-Teller transitions to the ground state of the final nucleus
and neutrinoless double-beta decay nuclear matrix elements.
The correlation seems robust across $pf$-shell nuclei.
\end{abstract}

\section{Introduction}
Neutrinoless double-beta ($0\nu\beta\beta$) decay
is possibly the best probe available
to test whether neutrinos are its own antiparticle.
While present $0\nu\beta\beta$ decay half-life limits are impressive,
exceeding $10^{25}$~years for $^{136}$Xe, $^{76}$Ge and $^{130}$Te~\cite{KamLAND-Zen16,GERDA17,EXO17,CUORE17},
observing the decay rates predicted by
the standard model of particle physics 
demands improving the half-life sensitivity of present experiments
by two orders of magnitude,
if the ordering of the neutrino masses is ``inverted",
or by four orders of magnitude, if it is ``normal"~\cite{agostini_discovery}.
The estimated sensitivities, however, depend critically in the 
nuclear matrix elements (NMEs) that control the decay, $M^{0\nu\beta\beta}$~\cite{engelmen_review}
\begin{equation}
\label{eq:half-life}
[T^{0\nu}_{1/2}]^{-1}=G^{0\nu}
\left|M^{0\nu\beta\beta}\right|^2  m_{\beta\beta}^2\,,
\end{equation}
with $G_{0\nu}$ a phase-space factor,
and $m_{\beta\beta}=\sum_{l} U_{el} m_l/m_e$
a combination of the neutrino masses and mixing matrix,
normalized to the electron mass.
Strategies to design future experiments
exploring ``inverted" and ``normal" scenarios
require a good knowledge of the NMEs.

Unfortunately nuclear theory is currently not in a position
to provide very reliable NMEs.
The best available calculations disagree between each other
by a factor two or three
when comparing results obtained with
different nuclear many-body approaches~\cite{engelmen_review}.
These calculations are tested against experimental data on several
nuclear structure properties:
excitation spectra, electromagnetic and $\beta$ decays~\cite{cau05},
knockout and transfer reactions~\cite{freeman-12,brown-14},
$\beta\beta$ decay with the emission of two neutrinos~\cite{barabash-15}...
Various many-body approaches generally find good agreement with data
when confronting these properties.
By contrast,
they differ significantly on their prediction of $0\nu\beta\beta$ decay NMEs.
So far no nuclear observable has been found to be especially well correlated
to $0\nu\beta\beta$ decay.
The observation that $0\nu\beta\beta$ decay seems to be subtle on its own way
has prevented to pin down the value of the NMEs.

\section{Double Gamow-Teller transitions and $\beta\beta$ decay}
Gamow-Teller (GT) strength distributions
measured via charge-exchange experiments~\cite{ichimura-06,frekers-13,yako-ca48}
are good tests of the many-body calculations.
GT distributions have the advantage that they can probe transitions
not accessed in $\beta$ decay due to the limited $Q_{\beta}$ value,
and in addition they exhibit collective behavior
best manifested in the GT giant resonance.
Recently, there is a strong experimental interest
in performing double charge-exchange experiments
to access the novel nuclear collectivity
indicated by double GT (DGT) transitions~\cite{cappuzzello,takaki-aris,uesaka-nppac,takahisa-17}.
Similarly to the GT case, this brings the opportunity
to assess the validity of many-body calculations of $\beta\beta$ decays
by testing predictions against measured DGT distributions. 

Double charge-exchange reactions are a result of the strong interaction.
On the other hand $\beta\beta$ decay
is governed by the weak interaction---the reason why it is related to neutrinos.
Nonetheless it is important to keep in mind that in both cases
a pair of neutrons can become a pair of protons,
via nucleon exchange in charge-exchange reactions,
or via nucleon decay in $\beta\beta$ decay.
In addition, the two processes are sensitive to the spin of the nucleons.
We also note that $0\nu\beta\beta$ decay
occurs at second-order in the weak interaction,
and in principle it should depend explicitly on the states
of the intermediate nucleus.
However, it does not make much difference
if one uses the so-called ``closure approximation"
that brings the NME
only dependent on the initial $(i)$ and final $(f)$ nuclear states~\cite{horoi-nonclosure-48ca}.
When doing so
the $0\nu\beta\beta$ decay and DGT matrix elements read~\cite{engelmen_review,2GT0nbb}
\begin{eqnarray}
&M^{0\nu\beta\beta}(i\rightarrow f)&
=M_{GT}^{0\nu}+\frac{M_F^{0\nu}}{g_A^2}+M_T^{0\nu}= \sum_{X=GT,F,T}
\bra{f} \sum_{a,b} H_X(r_{ab})\,S_X\, \tau^+_a \tau^+_b \ket{i}\,,\\
\label{eq:2GT}
&M^{DGT}(i\rightarrow f)&
=\bra{f} \sum_{a,b}
[{\bm \sigma}_a \tau^+_a \times {\bm \sigma}_b \tau^+_b]^{\lambda} \ket{i}\,,
\end{eqnarray}
where the sums run over all nucleons, $\tau$ represents the isospin,
and the DGT operator is coupled to $\lambda=0,2$.
$S_X$ denote the GT, Fermi and tensor spin structures
that contribute to $0\nu\beta\beta$ decay,
and $H_X(r_{ab})$ are the so-called neutrino potentials
corresponding to each spin structure
---divided by $g_A^2$ in the Fermi case.
The neutrino potentials depend on the distance between the decaying nucleons, and are given by~\cite{engelmen_review}
\begin{eqnarray}
\label{eq:nu_pot}
H_X(r_{ab})
= \frac{2R}{\pi} \int_0^\infty \!\!\! q^2 \, dq
\frac{j_X(q\,r_{ab}) h_{X}(q)}
{q\left(q+\mu\right)}\,,
\end{eqnarray}
with ${\bm q}$ the momentum transfer,
$R=1.2A^{1/3}$~fm the nuclear radius,
$j_F(q\,r_{ab})=j_{GT}(q\,r_{ab})=j_0(q\,r_{ab})$
and $j_T(q\,r_{ab})=j_2(q\,r_{ab})$ spherical Bessel functions,
and $\mu$ the so-called closure energy~\cite{horoi-nonclosure-48ca}.
The explicit form of the $h_{X}(q)$ functions
can be found in Ref.~\cite{menendez_heavy}.

The $M_{GT}^{0\nu}$ part with $S_{GT}={\bm \sigma}_a \cdot {\bm \sigma}_b$
is by far dominant in $0\nu\beta\beta$ decay,
accounting for almost 90\% of the total NME.
Therefore the operator structure that mainly drives $0\nu\beta\beta$ decay
is that of DGT transitions with $\lambda=0$ coupling. 
Overall, the main differences between the two processes
are the presence of the neutrino potentials in $0\nu\beta\beta$ decay,
and the capability of DGT transitions
to connect states that differ in up to two units of angular momentum.
In addition, in practice $0\nu\beta\beta$ NMEs
only refer to decays to the ground state of the final nucleus.
Other decays are very suppressed, or energy forbidden,
due to the small---about a couple of MeV at most---$Q_{\beta\beta}$ values.
In contrast, DGT transitions can populate final states
excited up to several tens of MeV.

\section{Results for DGT and $0\nu\beta\beta$ decay matrix elements}
The entire DGT distribution of the $\beta\beta$ emitter $^{48}$Ca
was recently calculated in Ref.~\cite{2GT0nbb},
using the shell model many-body framework.
Ref.~\cite{2GT0nbb} considered a typical shell model space
consisting of the $pf$-shell---this is,
including the 0$f_{7/2}$, 1$p_{3/2}$, 1$p_{1/2}$ and 0$f_{5/2}$
single-particle orbitals---and
an extended one comprising the $sd-pf$ configuration space---with
the 0$d_{5/2}$, 1$s_{1/2}$ and 1$d_{3/2}$ orbitals included as well.
The results predicted a DGT distribution relatively independent
of the size of the configuration space and the shell model interaction used.

\begin{figure}[t]
	\begin{center}
		\includegraphics[width=.9\textwidth]{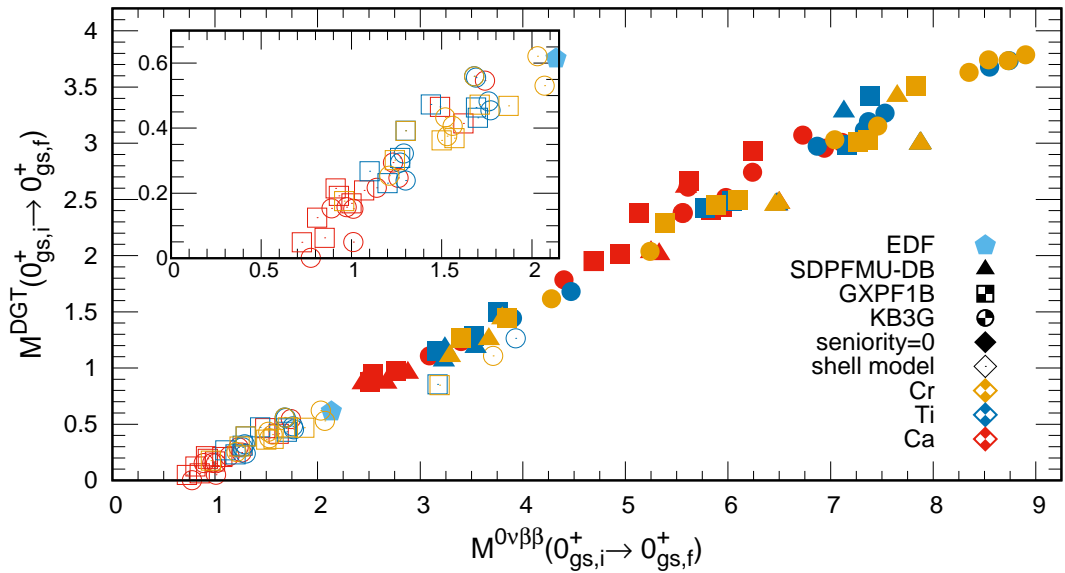}	
	\end{center}	
	\caption{\label{fig:nme}	
		Correlation between
		the $0\nu\beta\beta$ decay NME
		$M^{0\nu\beta\beta}(0^+_{gs,i} \rightarrow 0^+_{gs,f})$
		and the DGT transition to the ground state of the final nucleus
		$M^{\rm DGT}(0^+_{gs,i} \rightarrow 0^+_{gs,f})$.
		Shell model results for initial
		calcium (Ca), titanium (Ti) and chromium (Cr) isotopes
		are shown in red, blue and orange, respectively.
		Calculations with the $pf$-shell
		GXPF1B~\cite{gxpf1b}, KB3G~\cite{kb3g}
		and $sd-pf$ SDPFMU-DB~\cite{iwata-prl2016} interactions
		are denoted by squares, circles and triangles, respectively.
		Open symbols indicate fully correlated nuclear states,
		while filled ones refer to nuclei restricted to seniority-zero configurations. The shell model results are compared to
		the non-relativistic energy-density functional (EDF) theory
		$^{48}$Ca result from Ref.~\cite{rodriguez} (light blue pentagon).
		The inset zooms in the results corresponding to fully correlated states.
	}
\end{figure}

An analysis of the sensitivity of the total DGT strength distribution
to isovector---dominated by like-particle---and 
isoscalar---to which only proton-neutron contribute---pairing correlations
showed that isovector pairing correlations
tend to push the DGT giant resonance to higher energies~\cite{2GT0nbb}.
Combining this sensitivity with the well-known dependence
of the $0\nu\beta\beta$ decay NME
to isovector pairing~\cite{caurier-08}, Ref.~\cite{2GT0nbb} showed that,
for moderate increases of the isovector pairing correlations,
the energy of the $^{48}$Ca DGT giant resonance
and the $0\nu\beta\beta$ decay NME value follow a simple linear correlation.
The fact that a collective property such as the energy of the DGT giant resonance
might be related to the value of an individual matrix element
is very promising, and supports the similarity
between DGT transitions and $0\nu\beta\beta$ decay.
Further, the relation suggests that
nuclear structure measurements of DGT transitions
can be used to study, indirectly, $0\nu\beta\beta$ decay.

If there is a DGT transition
that can be expected to be especially related
to the $0\nu\beta\beta$ decay NME
this is the DGT transition to the ground state of the final nucleus.
In such case the final states
of the $0\nu\beta\beta$ decay and DGT transitions are the same.
Moreover, if the initial state is a $0^+$ state---as the ground states
of every even-even nucleus are--- the DGT transition to the ground state
can only be mediated by the $\lambda=0$ channel,
which implies that the spin structure of the DGT transition
and the (dominant part of) the $0\nu\beta\beta$ decay are common as well.
This implies that when only ground states are involved,
and taking into account the validity of the closure approximation,
the only difference between the $0\nu\beta\beta$ decay and the DGT transition
matrix elements
is caused by the neutrino potential $H(r_{ab})$.
The potential signals the fundamental difference between the two processes:
while $0\nu\beta\beta$ decay is driven by the weak interaction,
DGT transitions, explored in charge-exchange experiments,
are a result of the strong force.

Figure~\ref{fig:nme} shows the results of the shell model calculation
of $0\nu\beta\beta$ decay NMEs
compared to the DGT transitions to the ground state.
The decays of calcium (shown in red), titanium (blue)
and chromium (orange) isotopes are shown.
Figure~\ref{fig:nme} considers
a total of twenty six initial even-even nuclei, 
from the lightest $^{42}$Ca to the heaviest $^{60}$Cr.
In addition, Fig.~\ref{fig:nme} explores two kind of nuclear states:
the open symbols correspond to full shell model states,
while the filled symbols indicate simplified states
that entirely consist of pairs of neutrons and pairs of protons
coupled to $0^+$---known as seniority-zero states.
Finally, Fig.~\ref{fig:nme} shows results
obtained using different configuration spaces and shell-model interactions:
the $pf$-shell interactions GXPF1B~\cite{gxpf1b} (squares)
and KB3G~\cite{kb3g} (circles),
and the extended $sd-pf$ space
SDPFMU-DB interaction~\cite{iwata-prl2016} (triangles).
The results were obtained with the shell model codes KSHELL~\cite{kshell}
and NATHAN~\cite{cau05}.

Figure~\ref{fig:nme} highlights a very good linear correlation
between $0\nu\beta\beta$ decay NMEs and DGT transitions to the ground state.
The correlation seems robust.
It is valid across a very wide range of nuclei
in this region---mass numbers $42\leq A\leq60$---and it is also common
to the three shell model interactions used,
in the usual and extended configuration spaces.
Furthermore, the linear correlation remains valid
regardless of whether we consider fully correlated shell model states
or fairly simple seniority-zero wave functions.
Because seniority-zero nuclear states
overpredict the $0\nu\beta\beta$ decay NMEs,
the last property implies that the correlation is valid
for a very wide range of NME values $0\lesssim M^{0\nu\beta\beta}\lesssim 10$.
Such a broad range encompasses the factor-two-or-three range
of NME values that covers the results of
the different many-body calculations for any $\beta\beta$ emitter.

It is interesting to note that the $^{48}$Ca results obtained
in the framework of energy density functional theory~\cite{rodriguez}
(light blue pentagon in Fig.~\ref{fig:nme})
are consistent with the linear correlation found in the shell model calculations.
Reference~\cite{2GT0nbb} extended to heavier nuclei the comparison
of $0\nu\beta\beta$ decay and DGT transitions to ground states,
finding a linear correlation
very similar to the one in Fig.~\ref{fig:nme}~\cite{2GT0nbb}.
Again, for heavier nuclei the results from energy-density functional theory
agree well with the correlation observed in the shell model calculations.
In contrast, the results of the 
quasiparticle random phase approximation (known as QRPA)~\cite{simkovic-11}
behave very differently,
with DGT transition values that tend to be persistently small~\cite{2GT0nbb}.

How can we understand the existence of a rather universal linear correlation
between $0\nu\beta\beta$ decay and DGT matrix elements?
We can try to answer this question by studying
the matrix element distributions as a function of
the distance between decaying/transferred nucleons, and of the momentum transfer.
For $0\nu\beta\beta$ decay the normalized distributions are defined as
\begin{eqnarray}
C_{GT}(r)= \langle f | \sum_{ab} \delta(r-r_{ab}) \,H_{GT}(r_{ab})\,
{\bm \sigma}_a\cdot{\bm \sigma}_b \,\tau_a \tau_b | i \rangle / M_{GT}^{0\nu}\,, \\
\label{eq:density_p}
C_{GT}(p)= \langle f | \sum_{ab} \delta(p-q) \,\hat{H}_{GT}(q)\,
{\bm \sigma}_a \cdot {\bm \sigma}_b \,\tau_a \tau_b | i \rangle / M_{GT}^{0\nu}.
\end{eqnarray}
An analogous definition can be used for DGT transitions,
except that in that case there is no neutrino potential
and the spin structure is defined with a different normalization
$[{\bm \sigma}_a \times {\bm \sigma}_b]^0=
\frac{-1}{\sqrt{3}}{\bm \sigma}_a \cdot {\bm \sigma}_b$
(the different definition does not affect the normalized distributions).

\begin{figure}[t]
\begin{center}
	\includegraphics[width=.85\textwidth]{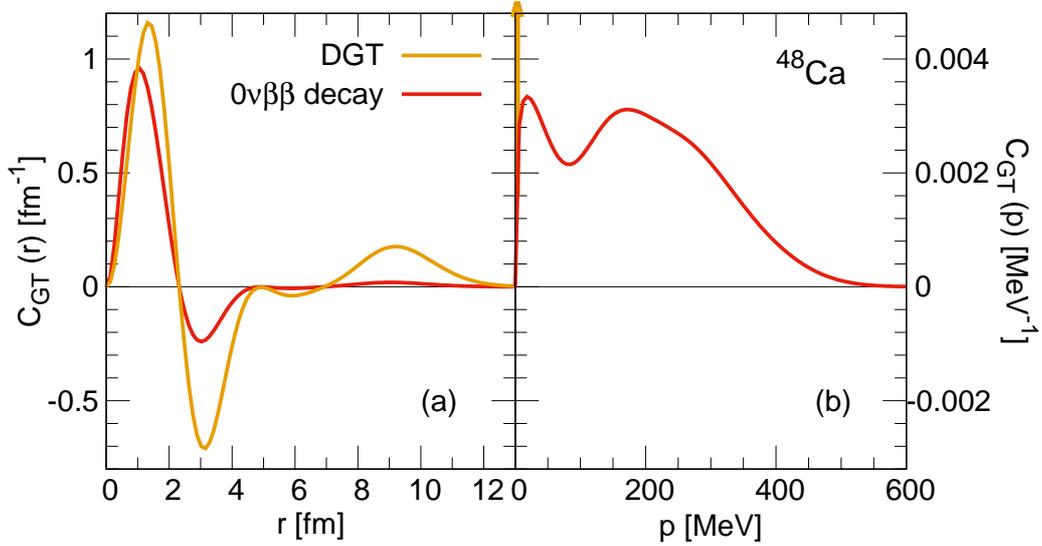}
\end{center}	
	\caption{\label{fig:density}
	Normalized distributions of the
	Gamow-Teller part of the $0\nu\beta\beta$ decay
	NME $M^{0\nu}_{\rm GT}$ (red) and
	and the DGT matrix element $M^{\rm DGT}$ (orange), 
	as a function of (a) the internucleon distance $C_{GT}(r)$,
	(b) the momentum transfer $C_{GT}(p)$.
	The results correspond to $^{48}\textrm{Ca}$
	calculated with the $pf$-shell GXPF1B interaction.}
\end{figure}

Figure~\ref{fig:density} shows the normalized radial and momentum
distributions for $^{48}$Ca.
The results show that the momentum distribution of the two processes
is very different:
while the DGT transitions in Eq.~(\ref{eq:2GT})
do not entail any momentum transfer\footnote{The identity that
replaces the neutrino potential in Eq.~(\ref{eq:density_p}) can be seen
as the Fourier transform of a delta function in momentum space.},
typical transferred momenta amount to $p\sim100$~MeV in $0\nu\beta\beta$ decay.
In contrast, the radial distributions of the two processes are much more similar:
they are dominated by short-range internucleon distances of $r\lesssim2$~fm.
Moreover, the longer-range contributions that contribute to DGT transitions
partially cancel, making the effective DGT range even more similar to that of $0\nu\beta\beta$ decay.
We have observed that this partial cancellation is common to all the other
DGT transitions in Fig.~\ref{fig:nme}.
We have to stress that
the very short-range character of $0\nu\beta\beta$ decay and DGT transitions
is a somewhat surprising outcome of the many-body calculations.
Its validity should be carefully checked in more controlled
ab initio calculations, such as those in Ref.~\cite{Pastore17b}.

The short-range character of $0\nu\beta\beta$ decay
and DGT transitions to the ground state
can explain the simple linear correlation observed in Fig.~\ref{fig:nme}.
References~\cite{bogner-10,bogner-12} showed that
when an operator probes only the short-range physics of low-energy states,
the corresponding matrix elements factorize
into a universal operator-dependent constant
times a state-dependent number common to all short-range operators.
Therefore, a linear relation is expected between two short-range operators,
such as $0\nu\beta\beta$ decay
and the corresponding DGT transition to the ground state.
This explanation is also consistent
with the quasiparticle random phase approximation results not being correlated:
for this many-body method $0\nu\beta\beta$ decay NMEs
are of short-range as in the shell model,
but DGT transitions receive important contributions from much longer
internucleon distances~\cite{simkovic-11,simkovic-08}.
To clarify this picture,
further work should explore the physics of the long-range contributions
in detail.

\section{Conclusions}
We have used the nuclear shell model
to study the similarity between $0\nu\beta\beta$ decay NMEs
and DGT transitions to the ground state,
for a collection of twenty six transitions in the $pf$-shell.
We have found that the $0\nu\beta\beta$ decay and DGT matrix elements
are very well correlated.
The correlation is common to all the transitions studied,
and does not depend on the details of the nuclear interaction used,
or the nuclear structure correlations
permitted in the initial and final nuclei.
Moreover, a result for $^{48}$Ca obtained with energy-density functional theory
is consistent with the correlation found in the shell model calculations.
Our findings suggest that future measurements of DGT transitions
in double charge-exchange experiments
can provide very valuable insights on $0\nu\beta\beta$ decay.


\section*{Acknowledgments}
This work was supported by JSPS KAKENHI Grant (25870168, 17K05433) and 
CNS-RIKEN joint project for large-scale nuclear structure calculations.
It was also supported by MEXT and JICFuS as a priority issue 
(Elucidation of the fundamental laws and evolution of the universe, hp170230) 
to be tackled by using Post K Computer.


\section*{References}
\begin{thebibliography}{9}

\bibitem{KamLAND-Zen16}
A. Gando {\em et~al.\/} (KamLAND-Zen Collaboration), {\em Phys. Rev.
	Lett.\/} {\bf 117} 082503 (2016).

\bibitem{GERDA17}
M. Agostini {\em et~al.\/} (GERDA Collaboration), {\em Nature\/} {\bf 544}
47 (2017).

\bibitem{EXO17}
J. B. Albert {\em et~al.\/} (EXO Collaboration), arXiv:1707:08707.

\bibitem{CUORE17}
C. Alduino {\em et~al.\/} (CUORE Collaboration), arXiv:1710:07988.

\bibitem{agostini_discovery}
M. Agostini, G. Benato, and J. A. Detwiler, {\em Phys. Rev. D} {\bf 96} 053001 (2017).

\bibitem{engelmen_review}
J. Engel and J. Men{\'e}ndez, {\em Rep. Prog. Phys.} {\bf 45} 014003 (2017).

\bibitem{cau05}
E. Caurier, G. Mart{\'i}nez-Pinedo, F. Nowacki, A. Poves and A.~P. Zuker {\em
	Rev. Mod. Phys.\/} {\bf 77} 427 (2005).

\bibitem{freeman-12}
S. J. Freeman, and J. P. Schiffer {\it et al.},
{\em J. Phys. G: Nucl. Part. Phys.} \textbf{29}, 124004 (2012).
            
\bibitem{brown-14}
B. A. Brown, M. Horoi, and R. A. Sen'kov, 
{\em Phys. Rev. Lett.} \textbf{113}, 262501 (2014).
      
\bibitem{barabash-15} 
A. S. Barabash, 
{\em Nucl. Phys. A} \textbf{935}, 52 (2015).

\bibitem{ichimura-06} 
M. Ichimura, H. Sakai, and T. Wakasa, 
{\em Prog. Part. Nucl. Phys.} \textbf{56}, 446 (2006).
      
\bibitem{frekers-13} 
D. Frekers, P. Puppe, J. H. Thies, and H. Ejiri, 
{\em Nucl. Phys. A} \textbf{916}, 219 (2013).

\bibitem{yako-ca48}
K. Yako {\it et al.}, 
{\em Phys. Rev. Lett.} \textbf{103}, 012503 (2009).


\bibitem{cappuzzello} F. Cappuzzello, M. Cavallaro, C. Agodi, 
M. Bondi, D. Carbone, A. Cunsolo and A. Foti,
{\em Eur. Phys. J. A} \textbf{51}, 145 (2015).

\bibitem{takaki-aris} 
M. Takaki {\it et al.}, {\em JPS Conf. Proc.} \textbf{6}, 020038 (2015).

\bibitem{uesaka-nppac}
T. Uesaka {\it et al.}, 
RIKEN RIBF NP-PAC, NP1512-RIBF141 (2015).
      
\bibitem{takahisa-17}
K. Takahisa {\it et al.}, arXiv:1703.08264. 
        
\bibitem{horoi-nonclosure-48ca}
R. A. Sen'kov and M. Horoi, {\em Phys. Rev. C} \textbf{88}, 064312 (2013).

\bibitem{2GT0nbb}
N. Shimizu, J. Men{\'e}ndez, and K. Yako, arXiv:1709:01088.

\bibitem{menendez_heavy}
J. Men{\'e}ndez, {\em J. Phys. G. Nucl. Part. Phys.} {\bf 45} 014003 (2018).

\bibitem{caurier-08}
E. Caurier, J. Men\'endez, F. Nowacki, and A. Poves, 
Phys. Rev. Lett. \textbf{100}, 052503 (2008).     

\bibitem{gxpf1b}
M. Honma, T. Otsuka, and T. Mizusaki, {\em RIKEN Accel. Prog. Rep.} \textbf{41}, 32 (2008).
      
\bibitem{kb3g} 
A. Poves, J. Sanchez-Solano, E. Caurier, and F. Nowacki, 
{\em Nucl. Phys. A} \textbf{694}, 157 (2001).

\bibitem{iwata-prl2016}
Y. Iwata, N. Shimizu, T. Otsuka, Y. Utsuno, 
J. Men\'endez, M. Honma, and T. Abe, 
{\em Phys. Rev. Lett.} \textbf{116}, 112502 (2016).

\bibitem{kshell}
N. Shimizu, arXiv:1310.5431.

\bibitem{rodriguez} 
T. R. Rodr\'iguez, G. Mart\'inez-Pinedo, {\em Phys. Lett. B} \textbf{719}, 
174 (2013).

\bibitem{simkovic-11}
F. {\v S}imkovic, R. Hod\'ak, A. Faessler, and P. Vogel, 
{\em Phys. Rev. C} \textbf{83}, 015502 (2011).

\bibitem{Pastore17b}
S. Pastore, J. Carlson, V. Cirigliano, W. Dekens, E. Mereghetti and R.~B. Wiringa, arXiv:1710.05026.

\bibitem{bogner-10}
E. R. Anderson, S. K. Bogner, R. J. Furnstahl, and R. J. Perry, 
{\em Phys. Rev. C} \textbf{82}, 054001 (2010).
    
\bibitem{bogner-12}
S. K. Bogner, and D. Roscher,
{\em Phys. Rev. C} \textbf{86}, 064304 (2012).    

\bibitem{simkovic-08}
F. {\v S}imkovic,  A. Faessler, V. Rodin,P. Vogel, and J. Engel, 
{\em Phys. Rev. C} \textbf{77}, 045503 (2008). 


\end{thebibliography}
\end{document}